\documentstyle[12pt]{article}

\def\ref#1{$^{#1)}$}
\begin{document}
\begin{titlepage}
\begin{center}
%         \hfill    LBL-32491 \\
%          \hfill    UCB-PTH-xx/xx \\

\vskip .05in

{\large \bf Unification Yang-Mills Groups and Representations with CP as a
Gauge Symmetry}
\vskip .05in
Huazhong Zhang\footnote{e-mail: HZHANG@CCAIX.JSUMS.EDU, Present address:
P.O.Box 17660, Jackson State University, Jackson, MS 39217}\\[.1in]
{\it  Theoretical Division, T-8, Los Alamos National Laboratory,\\
Los Alamos, NM 87545}\footnote{Where work supported by a DOE fellowship}
\end{center}
%\vskip .5in
\begin{abstract}
%insert abstract here
We investigate more generally the possible unification Yang-Mills groups
$G_{YM}$ and representations with CP as a gauge symmetry. Besides the possible
Yang-Mills groups $E_8$, $E_7$, $SO(2n+1)$, $SO(4n)$, $SP(2n)$, $G_2$ or $F_4$
(or a product of them) which only allow self-contragredient
representations, we present other unification groups $G_{YM}$ and
representations which may allow CP as a gauge symmetry. These include
especially $SU(N)$ containing Weyl fermions and their CP-conjugates from
low-energy spectra in a basic irreducible representation (IR). Such an example
is the 496-dimensional basic IR (on antisymmetric tensors of rank two) of
$SU(32)$ containing $SO(32)$ as a subgroup in the adjoint IR, or $SU(248)$
in a fundamental IR containing $E_8$ as a subgroup in the adjoint IR.
Our consideration also leads to the construction of a physical operator (CP)
intrincially as an inner automorphism of order higher than two for the
unification group. We have also generalized the possible groups as unification
$G_{YM}$ to include non-semisimple Lie groups with CP arising as a gauge
symmetry. In this case with $U(1)$ ideals in the $G_{YM}$, we found that the
$U_Y(1)$ for weak hypercharge in the standard model or a $U(1)$ gauge symmetry
at low energies in general is traceless. Possible relevance to superstring
theory is also briefly discussed. We expect that our results may open new
alternatives for unified model building, especially with deeper or more
generalized understanding of anomaly-free theories.
\end{abstract}
\end{titlepage}
\newpage
\renewcommand{\thepage}{\arabic{page}}
\setcounter{page}{1}
%THIS IS PAGE 1 (INSERT TEXT OF REPORT HERE)
\section{Introduction}
     It is known that there is a CP asymmetry in particle physics as
observed in $K^0-\bar{K^0}$ system [1]. The CP asymmetry has been
widely studied related to CKM matrix of quark mixing in the standard model,
superweak interaction models, etc. Although the full determination of
the underlying mechanism for CP asymmetry is still under investigation,
it is nevertheless an experimental fact that CP violation exists in nature.

In the standard model, an effective $\bar{\theta}$ term can be added to the QCD
lagrangian due to instanton tunneling, where
\begin{equation}
\bar{\theta}={\theta}_{QCD} + arg\{det~ m_q\}~
\end{equation}
with $m_q$ being the quark mass matrix.
The experimental upper bound on the neutron-electric dipole moment (EDMN)
limit the strong CP violating parameter $\bar{\theta}$ to be only
$\bar{\theta}\le 2\times 10^{-10} (mod ~2\pi)$ [1]. However, the
(almost) vanishing of strong CP asymmetry is unnatural if we believe
in 't Hooft`s naturalness condition [2]
that {\it a parameter is only allowed to be very small if setting it
to zero increases the symmetry}. Since we know that CP is broken in nature, by
setting $\bar{\theta}$ to zero the symmetry does not increase. This is
the strong CP problem in particle physics.

There are several possible solutions to the strong CP problem as proposed.
A well-known solution [3] is to introduce an anomalous global $U(1)_{PQ}$
(Peccei and Quinn) symmetry in addition to the standard model. The $U(1)_{PQ}$
symmetry is realized nonlinearly at low energies, leading to a light
pseudo-Goldstone boson, the axion [4] whose coupling may be adjusted [5]
to be consistent with the observation.

The second possible solution is that the up quark (a light quark) is massless.
There is an anomalous chiral U(1) global symmetry at the QCD scale,
rendering the $\bar{\theta}$ unphysical, in order to eliminate the up quark's
Yukawa coupling to the Higgs boson.

There is also a proposed non-perturbative solution with colored magnetic
monopoles as given by the present author [6]. The non-perturbative solution
in ref.6 is associated with the generalization of the gauge orbit space
in ordinary gauge theories on the compactified space to that
(generalized gauge orbit space) on the space with a topologically
non-contractile spherical boundary
in the presence of magnetic monopoles. Especially, a gauge orbit subspace
(restricted gauge orbit space) with gauge potentials and well-defined
globally continuous gauge transformations restricted on the space boundary
is introduced and investigated. With this generalization, the $\bar{\theta}$
may be quantized nonperturbatively to be vanishing or in a form of
$\bar{\theta}=2\pi/n $ with n being the topological charges [6] of the
monopole, so that the $\bar{\theta}$ can take the special values
such as $\pm 2\pi$, or naturally very small values such as $\le 10^{-10}$
due to the topological charges of the colored monopoles in the universe.
Our quantization in ref.6 can be derived from two different approaches.
One is due to the quantization of flux (proportional to the vacuum angle
$\bar{\theta}$) in the (restricted) gauge orbit space to have a well-defined
wave functional in the entire space including the space boundary. The
other approach is due to the global constraints of Gauss'law in the generalized
gauge orbit space associated with the non-abelian electric charges
proportional to $\bar{\theta}$ in the presence of colored monopoles.
A consequence of the non-perturbative solution to the strong CP problem
in ref.6 is that the (almost) vanishing of CP violation in the strong
interaction may imply the open universe. Moreover, it is also noted that
[6] in the presence (colored) monopoles, the corresponding QCD may need
to be generalized to that with a non-associative algebra if one needs
to formulate the theory exactly due to the violation of the ordinary
Bianchi identity, although perturbatively the algebraic structure for the
theory may be regarded as associative at least to a good approximation
(The violation for the associativity could be exponentially small as the
distance from the monopoles increases.).

There are also other arguments for the absence of strong CP asymmetry [7]
(see also [8]). However, these in ref.7 may not be consistent with the
$U(1)_A$ problem as stressed in ref.9.

There is also a possible solution associated with grand unification theory,
in which one imposes an exact discrete CP symmetry which is spontaneously
broken. The parameter $\bar{\theta}$ is then finite and calculable, and in a
class of models can be as small as needed for consistency with the
experimental limit [10]. This is a solution which may allow the CP being
viewed as arising from a discrete gauge symmetry, as initially discussed
in ref.9.

Note that as emphasized in ref.9, neither the axion nor the massless up
quark solution is natural in the 't Hooft sense, since the global U(1)
symmetry in either case is only approximate (anomalous),
and the $\bar{\theta}$ may be too large in the Peccei-Quinn case by
U(1) violating instanton effects [11]. Furthermore, these two solutions with an
anomalous global U(1) symmetry certainly cannot be protected by
a gauge symmetry simply because the global U(1) symmetry is anomalous.
These two solutions therefore may be spoiled by the operators arising from
the Planck mass physics [9].
However, the spontaneously broken CP (or P) models may be viable when the
quantum gravity effects are considered since no additional global symmetry is
imposed.

As it has been argued, the quantum gravity effects from wormhole [12],
virtual black hole [13] or nonperturbative effects in string theory [14],
may lead to violation of global symmetries in the effective theory below the
Planck scale [15]. However, both continuous gauge symmetries and unbroken
discrete subgroups of gauge symmetries (or discrete gauge symmetries) are
regarded as preserved by Planck-scale physics, since they are violated
neither by wormholes nor blackholes [16,17]. An interesting aspect of
the discrete gauge symmetries is that a black hole can carry discrete
hairs [16]. With this consideration of
quantum gravity effects and the spontaneously broken CP symmetry models,
Choi, Kaplan and Nelson et al in ref.9
have proposed that four-dimensional CP may arise as
a discrete gauge symmetry in theories with dimensional compactification
from Minkowski dimensions 8k+1, 8k+2 or 8k+3. The CP remains an exact
discrete gauge symmetry below the compactification scale and is
spontaneously broken at much lower scale (below $10^9$ Gev),
with a $\bar{\theta}$ smaller
than $10^{-10}$ as suppressed by inverse power of the Planck mass $m_P$,
as consistent with the observed CP violation in the $K^0-\bar{K^0}$ system
and the experimental limit on EDMN. The consideration initiated in ref.9
provides a connection between the Yang-Mills group $G_{YM}$
at the Planck scale (As in ref.9, the compactification may be taken near to the
Planck mass, we will simply refer them as Planck scale or $G_{YM}$ scale
in our discussions for convenience) as well as the strong CP problem.

Write the continuous local symmetry G of a higher dimensional theory as
$G=G_L\times G_g\times G_{YM}$, where $G_L=spin(d-1,1)$, the d-dimensional
Lorentz group, $G_g$ are d-dimensional general coordinate transformations
with positive jacobian and $G_{YM}$ is the internal (unification) Yang-Mills
group. In ref.9, the four-dimensional CP transformation is given by the
product
\begin{equation}
CP=X_L X_g X_{YM}~,
\end{equation}
with $X_L, X_g,$ and $X_{YM}$ in $G_L, G_g$ and $G_{YM}$ respectively.
The $X_{YM}$ is an inner automorphism of the internal Yang-Mills
group $G_{YM}$.
The $X_L$ as a matrix is a direct sum of four-dimensional Minkowski metric
and a (d-4)-dimensional real matrix satisfying $K_LK_L^T=1$ and
$det(K_L)=-1$. The $X_g$ must be chosen to include a compactified
coordinate transformation $\theta_i\rightarrow -\theta_i$, where
the vector fields $\partial /\partial \theta_i$ generate the
Cartan subalgebra of the continuous isometry group of the compactified
dimensions, so that it reverses the orientations of both four-dimensional
Minkowski space and the compactified manifold.
Then [9], the transformation $CP=X_L X_g X_{YM}$ exchanges
all gauge charges of four-dimensional fields with gauge charges of
their antiparticles, including gauge interactions arising from the
isometries of the compactified space. This class of models with CP
as a gauge symmetry includes the case of ten-dimensional popular superstring
theories with gauge group $SO(32)$ or $E_8\times E_8$, where
four-dimensional CP can arise as a discrete local symmetry.

As pointed out in ref.9, in order for the CP transformation to be an
element of the local symmetry $G=G_L\times G_g\times G_{YM}$ of the
underlying d-dimensional ($d>4$) theory, we need that a four dimensional
low energy fermion be contained in the same irreducible representation
of the local symmetry group as their complex conjugates. Since the CP
operation transforms a Weyl fermion to its complex conjugate which is in a
different irreducible representation of the Lorentz group in four dimensions,
to realize the CP as a local symmetry, the underlying theory must be in
certain higher spacetime dimensions [9].
With the inner automorphism $X_L$ and the $X_g$ chosen as described above, we
would like to emphasize that {\it we only need to
ensure that $X_{YM}$ is an inner automorphism of the
$G_{YM}$ reversing the sign for the gauge charges of all the four-dimensional
low-energy fields contained in an irreducible representation of $G_{YM}$},
since this is the condition for the CP operator to
play the role of four-dimensional CP transformation in low energies [9],
and since the CP operator is an element of the full local symmetry group it
is a local (gauge) symmetry in the underlying d-dimensional theory.
In ref.9, the unification (internal) Yang-Mills group $G_{YM}$ is only
limited to the following groups and their products:
$E_8, E_7, G_2, F_4, SO(2n+1), SO(4n), SP(2n)$. These are the groups
which have no complex representations. More explicitly, every
representation of a group listed above is equivalent to its contragredient
representation by an inner automorphism, namely it is self-contragredient.
It is the purpose of the present paper
to generalize the unification Yang-Mills groups and representations
beyond the groups listed above with
CP as a gauge symmetry. Since the possibilities for the unification Yang-Mills
groups and representations are interesting in particle physics, we expect that
our consideration and generalization are significant and may be of
importance for the unified model building.

We will organize our following sections thus: In section 2,
we will present the groups SU(N), SO(4n+2) and $E_6$ as the unification
Yang-Mills groups $G_{YM}$ and their self-contragredient irreducible
representations. Then in section 3, we will describe how the
$SU(N)$ and $SO(4n+2)$ groups in some of their complex representations
may be constructed as the
the unification Yang-Mills group with CP as a gauge symmetry. The construction
of the corresponding inner automorphism $X_{YM}$ will also be described.
We note that our purpose is to present how the above idea can be realized.
Chiral gauge anomalies may be canceled by adding other fermion IR's etc. at
the relevant heavy scale. we will not generally address anomalies
in detail in the present paper although we will give
a brief relevant discussion about anomalies in section 4 in which the
possibilities with non-semisimple groups as $G_{YM}$ are also considered.
We like to refer to
the fact that many authors have searched and investigated local
anomaly-free (both in gravitation and Yang-Mills sectors) theories with
more relaxed conditions following or generalizing the Green-Schwarz
mechanism (see ref.14 for a review) and found many possibilities [18-19].
It is assumed that our constructions for the $G_{YM}$ may
possibly correspond to anomaly-free theories in some mechanisms of
anomaly cancellation, which may be with more relaxed conditions
such as in refs.[14,18-19] or with the further understanding of
anomalies in future theories.
We expect that at, least some of our constructions for the $G_{YM}$
may correspond to anomaly-free theories and may possibly be useful for
the unified model building.
Our conclusions will be summarized in section 5. An appendix for the
Branching rules of the basic IR's of $SU(N)$ to $SO(N)$, etc. needed
will also be given.
Before go into the next section, we note that we may refer the Weyl fermions
at low energies, Majorana-Weyl fermions in 8k+2 dimensions, and
Majorana fermions in 8k+1, 8k+3 dimensions with a suitable [9] $X_L$
for CP to be a gauge symmetry all simply as fermions in our discussions
for convenience, the implied meaning should be clear correspondingly.

\section{$SU(N), SO(4n+2)$ and $E_6$ or Their Products as Unification
Yang-Mills Groups in Self-contragredient Representations}

Let us denote by G a semisimple Lie group with g as a generic element of G,
and $\cal G$ its corresponding Lie algebra with x as a generic element of
$\cal G$. Let $\omega (G)$ be a linear representation of G (i.e,
a homomorphism of G into the group of non-degenerate linear transformations
of the space $R^{dim(\omega)}$, its corresponding linear representation
$\omega (\cal G)$ is a homomorphism of $\cal G$ into the Lie algebra of
all linear transformations of the space $R^{dim(\omega)}$,
and the $\bar{\omega}(g)=[\omega (g)^{T}]^{-1}, g\in G$
is the contragredient representation. Here, T is the transpose (in a
chosen basis). Then the representations of the corresponding Lie algebra
$\cal G$ are then related as given by
\begin{equation}
\bar{\omega }(x)=-[\omega (x)]^T, x\in {\cal G}.
\end{equation}
Self-contragredient representations have been studied in refs.20-24.
We will use mostly the ref.23 by Dynkin as well as refs.20 and 22.

A representation $\omega $ is called self-contragredient if there
exists an inner automorphism, namely a conjugation by an element
$S\in{\omega (G)}$, such that
$\bar{\omega}(g)=S^{-1}\omega (g)S$. In terms of the Lie algebra
elements x in the $\omega$ as hermitian matrices,
this can also be written as
\begin{equation}
\omega (x)=-S^{-1}[\omega (x)]^TS~.
\end{equation}
Let $\cal H$ be a given Cartan subalgebra of $\cal G$, then
all the $\omega (h), (h\in{\cal H})$ are diagonal matrices with respect to
a canonical basis of $\omega$ (having a non-zero vector from each
weight subspace with the higher weight vector numbered first).
Let $\Pi=\{{\alpha}_i\mid i=1,2,...,n=rank(G)\}$
be the simple root system of $\cal G$ corresponding to the Cartan subalgebra
$\cal H$, and $e_{\alpha}$ be the root vector in the $\cal G$ corresponding
to a simple root $\alpha \in \Pi$. An automorphism of a Lie algebra is an
automorphism of the vector space of the algebra preserving the Lie algebraic
commutators. A mapping of the simple root system $\Pi$ with
$\alpha_i\rightarrow \alpha_{i'}$ onto itself preserving the Cartan matrix
$A_{ij}=2<\alpha_i,\alpha_j>/<\alpha_i,\alpha_i>$, is an automorphism of
$\cal G$ defined by
\begin{equation}
f(\alpha_i)=\alpha_{i'},f(e_{\alpha_i})=e_{\alpha_{i'}}~.
\end{equation}
For example, to every isometric mapping $\alpha_i\rightarrow \alpha_{i'}$
for the simple root system $\Pi$, the f defined by the above equation
is an automorphism of the $\cal G$. Obviously, it is also an automorphism
of the Dynkin diagram of $\cal G$.
It is known that (see theorem 0.2 in ref.23), each automorphism u of $\cal G$
can be uniquely written as the combination of an automorphism f as defined in
the above equation corresponding to an isometric mapping of the
simple root system (or f = automorphism of the Dynkin diagram) and an
inner automorphism $u_0$, i.e.
\begin{equation}
u=u_0f~.
\end{equation}
Therefore, an automorphism is inner if the corresponding f acts as the
identity on the Dynkin diagram.

Now for an irreducible representation, we have
\begin{equation}
\bar{\omega} (h)=-\omega(h),~h\in {\cal H}
\end{equation}
for the diagonal matrices $\omega (h)$. Hence it is clear that the weight
systems $\Delta$ of $\omega$ and $\bar{\omega}$ are linked by the relation
\begin{equation}
\Delta_{\bar{\omega}}=-\Delta_{\omega}~.
\end{equation}
Since what we are looking for in our consideration of CP as a gauge symmetry
is the $\cal G$ and $\omega$ for which there exists an
inner automorphism to transform the IR $\omega$ into its contragredient
IR $\bar{\omega}$, the above equation shows clearly that this is equivalent to
having an inner automorphism to reverse the signs of all the weights
(Yang-Mills charges). Thus for a self-contragredient IR $\omega$,
if $\lambda\in\Delta_{\omega}$ is a weight, then so is $-\lambda$.
The corresponding group G in the IR $\omega$ is then a
possibility for the unification Yang-Mills group $G_{YM}$ with CP
as a gauge symmetry.

Since it is known that [20] an IR $\omega$
is self-contragredient if and only if
the highest and lowest weights $\Lambda$ and $\Lambda_l$ in
its weight system $\Delta_{\omega}$ differ only by a sign, i.e.
$\Lambda_l=-\Lambda$, or an element of the Weyl group will transform
the $\Lambda$ to $-\Lambda$.
The general algorithm [23] for deriving the $\Delta_{\omega}$ shows that
for all the IR's of $SO(4n)$, we have $\Lambda_l=-\Lambda$. Therefore,
all the IR's of $SO(4n)$ are self-contragredient.
To consider the other groups, let us consider the operation $\theta$
defined by
\begin{eqnarray}
\theta e_{\pm\alpha_i}=e_{\mp\alpha_i},~\alpha_i\in \Pi~ .
\end{eqnarray}
{}From the commutation relation written as
$[e_{\alpha}, e_{-\alpha}]=<e_{\alpha},e_{-\alpha}>\alpha$, it follows that
\begin{equation}
\theta\alpha_i=-\alpha_i,~\alpha_i\in \Pi ~.
\end{equation}
Then, the $\theta$ obviously preserves the Cartan matrix and is an
automorphism which transforms an IR $\omega$ into its
contragredient $\bar{\omega}$, i.e.
\begin{equation}
\bar{\omega}({\cal G})=\omega(\theta{\cal G}) ~.
\end{equation}
Thus, we have
$\theta=u_0f_{\theta}$ with $f_{\theta}$ being the corresponding
automorphism of the Dynkin diagram. Since the inner automorphism $u_0$
by definition corresponds to an equivalence relation between the
representations $\omega({\cal G})$ and $\omega(u_0{\cal G})$, we have
the equivalence relation
\begin{equation}
\bar{\omega}({\cal G})\simeq\omega(f_{\theta}{\cal G}) ~.
\end{equation}
{\it Therefore, the IR $\omega (\cal G)$ is self-contragredient if and only if
the $\theta$ is an inner automorphism, or equivalently if and only if
the $f_{\theta}$ is an identity automorphism of the Dynkin diagram}.

According to Dynkin (see theorem 0.16 in ref.23), in order that $\theta$
be an inner automorphism of a semisimple Lie algebra $\cal G$ it is necessary
and sufficient that it be an inner automorphism of every simple ideal of
$\cal G$, we only need to consider the simple Lie algebras. From the
Dynkin diagrams of simple Lie algebras, one can easily see that [23] the
Dynkin diagrams of $E_8, E_7, G_2, F_4, SO(2n+1)$, and $SP(2n)~
(SP(2)\simeq SU(2))$ can only have identity automorphism, this implies
in particular that the $f_{\theta}$ must be identity, or equivalently the
$\theta$ must be an inner automorphism. The IR's for
these groups and their products are all self-contragredient. For $SO(4n)$,
as we have indicated, all the IR's are also self-contragredient, the
$\theta$ in fact must be also an inner automorphism
(see theorem 0.16 in ref.23). Therefore, we have seen that the IR's of
\begin{equation}
E_8, E_7, G_2, F_4, SO(2n+1), SO(4n), ~and~SP(2n)
\end{equation}
(or a product of them) are all self-contragredient.
These groups containing at least a family of low energy fermions
(may include a right-handed neutrino) as well as their antiparticles
in a IR, may be a possibility of the unification Yang-Mills group. The three
(or more) families may be in the direct sum of the three (or more)
IR's, or in one large enough IR as in superstring theory [14] with
the Yang-Mills group $E_8\times E_8$. These are the groups mentioned in the
ref.9 initially for the consideration of the CP as a gauge symmetry.
Obviously, the group element S for the inner automorphism $\theta$ corresponds
to the $X_{YM}$ in the CP transformation $CP=X_L X_g X_{YM}$ described earlier.
We note that for those groups, the inner automorphism $\theta$ by definition is
automatically of order two, i.e, the ${\theta}^2$ is an identity
transformation.

In the present paper, we are especially interested in the groups beyond those
listed above. From the Dynkin diagrams of $SU(n+1)~(n\ge 2), SO(4n+2)~(n\ge 2),
E_6, (SU(4)\simeq SO(6))$, one can easily see that they all can have an
automorphism which may be generally non-trivial and then must be necessarily
the $f_{\theta}$. We can then write
for $SU(n+1)~(n\ge 2)$,
\begin{equation}
f_{\theta}(\alpha_i)=\alpha_{n+1-i}~(i=1,2,..,n)~,
\end{equation}
for $SO(4n+2)~(n\ge 2)$,
\begin{eqnarray}
f_{\theta}(\alpha_i)=\alpha_i, i\le 2n-1~,\\
f_{\theta}(\alpha_{2n})=\alpha_{2n+1}, f_{\theta}(\alpha_{2n+1})=\alpha_{2n}~,
\end{eqnarray}
and for $E_6$,
\begin{equation}
f_{\theta}(\alpha_i)=\alpha_{6-i},(i\le 5), f_{\theta}(\alpha_6)=\alpha_6~.
\end{equation}
Thus, the $f_{\theta}$ in the above equations acts as a permutation of the
root system leaving the Dynkin diagram invariant. We assume the Dynkin
diagrams and simple roots are in the usual convention [25],
for example, the simple roots 1-5
correspond to the dots on the same line for $E_6$. For $SO(4n+2)$, the
simple roots 1-(2n-1) correspond to the dots on the same line and the
simple roots (2n) and (2n+1) are only connected to the root (2n-1).
For all the Dynkin diagrams for the simple Lie algebras, see table 5 in ref.25.
Since the highest weight $\Lambda$ determines an IR $\omega$,
a general theorem then follows.\\
\underline{{\it Theorem 1}}.
{\it An IR $\omega$ for a simple Lie algebra is
self-contragredient if and only if the transformation $f_{\theta}$
leaves the highest weight of $\omega$ invariant}.\\
To be explicit, the highest weight $\Lambda$ can be written as
\begin{equation}
\Lambda=m_1\lambda_1+m_2\lambda_2+...+m_r\lambda_r~,
\end{equation}
here $r=rank(G)$, and $r=n, 2n+1, and~ 6~for~SU(n+1), SO(4n+2), and~ E_6$
respectively, the $\lambda_i$ are the fundamental weights given by
\begin{equation}
\frac{2<\lambda_i,\alpha_j>}{<\alpha_j,\alpha_j>}=\delta_{ij}~~(i,j=1,2,...,r).
\end{equation}
As it is known that (see for example, theorem 0.9 in ref.23) the $\Lambda$ is
the highest weight of an IR $\omega$ of an $\cal G$ it is necessary and
sufficient that the numbers $m_i$ which can be written as
\begin{equation}
m_i=\frac{2<\Lambda,\alpha_i>}{<\alpha_i,\alpha_i>}~~(i=1,2,...,r),
\end{equation}
are all non-negative integers. Since the fundamental weights for a simple
Lie algebra form a (linearly independent) basis in the vector space
for the root system, the theorem 1 above can be also stated as\\
\underline{{\it Theorem 2}}. {\it An IR $\omega$ of the highest weight
$\Lambda=\Sigma~m_i\lambda_i$
for a simple Lie algebra is self-contragredient
if and only if the transformation $f_{\theta}$ on the simple root system
leaves all the non-negative integers $m_i$ invariant}.\\
Using this theorem for the self-contragredient condition
\begin{eqnarray}
f_{\theta}(m_i)=m_i=\frac{2<\Lambda,\alpha_{i'}>}
{<\alpha_{i'},\alpha_{i'}>},\\
\alpha_{i'}=f_{\theta}(\alpha_i),~~(i=1,2,...,r).
\end{eqnarray}
and the transformation $f_{\theta}$ in eqs.(12-15) for the
$SU(n+1), SO(4n+2)$, and $E_6$, the self-contragredient IR's for
these groups can then be easily listed as follows by the relations for
the $m_i$ for the highest weight $\Lambda$:\\
\underline{$\bf{SU(n+1)}$}:\\
IR's of $\Lambda=\Sigma~m_i\lambda_i$ such that
\begin{equation}
m_i=m_{n+1-i}~~(i=1,2,...,n).
\end{equation}
\underline{$\bf{SO(4n+2)}$}:\\
IR's of $\Lambda=\Sigma~m_i\lambda_i$ such that
\begin{equation}
m_{2n}=m_{2n+1}~.
\end{equation}
\underline{$\bf{E_6}$}:\\
IR's of $\Lambda=\Sigma~m_i\lambda_i$ such that
\begin{equation}
m_{1}=m_{5},~m_{2}=m_{4}~.
\end{equation}
We note in particular that the IR's of the highest weight
$\Lambda=\lambda_1+\lambda_n~for~SU(n+1)$, $\Lambda=\lambda_2~for~SO(4n+2)$,
and $\Lambda=\lambda_6~for~E_6$ of dimensions $d=n(n+2), (2n+1)(4n+1),$ and 78
respectively are the adjoint representations which are all real.
Therefore, the $SU(n+1), SO(4n+2), E_6$ (or a product of them) in the
self-contragredient IR's listed above may possibly be the unification
Yang-Mills group with CP as a gauge symmetry. For those self-contragredient
IR's, the $f_{\theta}$ leaves the highest weight invariant and reduces the
automorphism $\theta$ to an inner automorphism [23,21]. The corresponding
group element S for the inner automorphism $\theta$ then corresponds to
the $X_{YM}$ in the CP operation as a local symmetry. The automorphism
$\theta$ by definition is of order two, the same as that for the groups
$E_8, E_7, G_2, F_4, SO(2n+1)$, and $SP(2n)$ (or a product of them).
We note also that the inner automorphism S in eq.(4) by definition
must be either symmetric or antisymmetric as one can see easily.

In summary of this section, the unification Yang-Mills group $G_{YM}$
with CP as a gauge symmetry
may be in general a semisimple group, i.e. a simple group
$(E_8, E_7, G_2, F_4, SO(2n+1), SP(2n), SU(n+1), SO(4n+2), E_6)$
or a product of them with the groups $SU(n+1), SO(4n+2)$, and $E_6$
in a self-contragredient representation as satisfying the conditions
in eq.(23-25). This has then generalized the possibilities in ref.9 for
the $G_{YM}$ as a semisimple group with each simple ideal restricted to
that listed in eq.(13).
\section{SU(N) and SO(4n+2) in Complex Representations as the Unification
Yang-Mills Groups}
So far we have only considered the self-contragredient representations
for a semisimple Lie group. As clarified in the section 1,
to have CP as a local symmetry, a low energy
Weyl fermion and its complex conjugate need to be contained in the
same IR of the unification Yang-Mills group $G_{YM}$ [9]. The easiest way
is by simply putting all the low energy fermions and their complex
conjugates in an IR of the $G_{YM}$ as listed in eq.(13) or a product of
them, this is the possibility for the $G_{YM}$ discussed in ref.9. The simple
ideals for $G_{YM}$ has been generalized in the section 2 to include also
$SU(n+1), SO(4n+2)$, and $E_6$ in their self-contragredient representations.
One then may wonder if we can choose $G_{YM}$ in a complex representation
with CP as a gauge symmetry. For simple Lie groups, only $SU(n+1), SO(4n+2)$,
and $E_6$ can have complex representations as we have seen.
In this section, we will show explicitly that we may choose $SU(n+1)$ as
$G_{YM}$ in some complex IR's with CP as a gauge symmetry.

Our idea to show the possibility of a complex representation for the
$G_{YM}$ with CP as a gauge symmetry is due to the following observation:
Since the CP transforms a Weyl fermion to its antiparticle at low energies,
then to realize the CP as a gauge symmetry at the Planck scale and
spontaneously broken at much lower energy, an arbitrary Weyl fermion and its
antifermion in the low energy particle spectra as a fermion pair
need to be embedded into the same IR $\omega (G_{YM})$
of the unification Yang-Mills group
$G_{YM}$ with an inner automorphism $S=X_{YM}$ which exchanges the two
Weyl fermions in the pair. Since the $X_{YM}$ for the inner automorphism
is an element of the Yang-Mills group $G_{YM}$, it is of course a
gauge symmetry, and then the $CP=X_L X_g X_{YM}$ as clarified in the
section 1 is a local symmetry. Now if the IR $\omega$ is complex,
then the inner automorphism with the group element S cannot reverse the
sign of the Yang-Mills charges for all the fermions in the IR $\omega$,
but it must exchange each fermion pair which correspond to a
fermion-antifermion pair at the low energy spectra. Obviously,
two types of possibilities may follow:\\
(1). The fermion pairs in the $\omega$ with low-energy correspondence to
(Weyl) fermion-antifermion pairs do not have to be fermion-antifermion pairs
at the Planck scale. In this case, the inner automorphism $X_{YM}$ can
only reverse the signs for some of the Yang-Mills charges
(including all those corresponding to low energy gauge symmetries)
for the Weyl fermions in the fermion pairs, the other Yang-Mills
charges (corresponding to the Yang-Mills symmetries broken above low
energy) may then be either invariant or transformed more non-trivially
than a sign-reversing.\\
(2). The fermion pairs in the $\omega$ with low-energy correspondence to
(Weyl) fermion-antifermion pairs are also fermion-antifermion pairs
at the Planck scale. The inner automorphism $X_{YM}$ may
reverse the signs of all the Yang-Mills charges for the Weyl fermions
in such pairs, but not for all the additional fermions needed to form
the IR $\omega$.

The central idea above then may be summarized as the follows:
To have the four-dimensional $CP$ arise as a gauge symmetry, we only
need to ensure that the a local symmetry $CP=X_L X_g X_{YM}$ at the Planck
scale induce CP transformations in four-dimensions to exchange Weyl fermion
and its antifermion at low energies. However, at the Planck scale,
such fermion pairs corresponding to fermion-antifermion pairs at
low energies need to be exchanged within each pair by the local
CP, the fermions in each pair do not need to be a fermion-antifermion
pair or there are additional fermions whose Yang Mills charges do not
all change their signs under the local CP. Note that in the (1),
there are in general additional fermions at the Planck Scale
other than those with correspondence to fermions at low energies.
Of course, the additional fermions need to obtain their masses at higher
energies, and will then not be contained in the low energy spectra.

We will realize possibilities in (1) described above. Our method
to construct such a complex IR $\omega$ for the $G_{YM}$ is to
embed a relevant self-contragredient IR $\omega_0$ for a symplectic or
orthogonal group into it. To do so, let us first recall some more facts
for self-contragredient representations needed to clarify our construction.
We will provide what we need in a theorem [20, 23] and some remarks,
for details, see the ref.[20-24].

\underline{{\it Theorem 3}}.{\it (Theorems 5,6 in [20], and [23,22]).
A representation $\omega$
for a semisimple Lie group G (algebra $\cal G$) is self-contragredient
if and only if there exists an invariant bilinear form $Q(\xi,\eta)$
under the group transformations in its representation vector space V.
If the representation $\omega$ is irreducible, then the bilinear form
$Q(\xi,\eta)$ is either symmetric, i.e. $Q(\xi,\eta)=Q(\eta,\xi)$, or
skew-symmetric, i.e. $Q(\xi,\eta)=-Q(\eta,\xi)$, for any $\xi,\eta\in V$,
the bilinear form $Q(\xi,\eta)$ is furthermore
nondegenerate and unique up to a constant factor. The symmetric
(skew-symmetric) bilinear form $Q(\xi,\eta)$ implies that the
IR $\omega$ is equivalent to a subgroup of proper-orthogonal (symplectic)
transformations on the vector space V.}

\underline{{\it Remark 1}}. The IR $\omega$ with symmetric (skew-symmetric)
invariant bilinear form $Q(\xi,\eta)$ in the representation vector space
V is called an orthogonal or real (symplectic or pseudoreal) representation,
as it is equivalent to
a representation with proper-orthogonal (symplectic) matrices for the group.
More precisely, by the theorem 3, the orthogonal (symplectic) IR $\omega$
is equivalent to a subgroup of $SO(d)$ ($SP(d)$) in the fundamental vector
representation, where $d=d(\omega)$ denotes the dimension of $\omega$ and
it must be even in the symplectic case.

\underline{{\it Remark 2}}.
It is known also that the automorphism matrix S in
eq.(4) for an IR $\omega$ with a symmetric (skew-symmetric) invariant
bilinear form is symmetric (skew-symmetric), and unique up to a constant.
The S as a matrix in the $\omega (G)$ must then be proper-orthogonal
(symplectic) when the $\omega (G)$ elements are all in terms of
proper-orthogonal (symplectic) matrices. In particular, the symmetric matrix
S which is also orthogonal must then be the unit matrix. In this case, the
S represents an identity automorphism as in eq.(4), and all the
matrices for the group elements are real. In the case of skew-symmetric
S for the IR $\omega$, a symplectic and skew-symmetric matrix cannot be
equivalent to the unix matrix and the inner automorphism must be non-trivial,
i.e. the matrices for the group elements cannot be all real although the
$\omega$ and its contragredient IR are equivalent. This clarifies the
self-contragredient IR in the meaning of "real" and "pseudoreal".
We will now give our discussions about the $SU(N)$ or $SO(4n+2)$ in
complex representations as $G_{YM}$ respectively.

\begin{center}
{\bf I. $\bf{G_{YM}=SU(N)}$}
\end{center}

In section 2, we know that semisimple Lie groups in their
self-contragredient IR's could be the $G_{YM}$
with CP as a gauge symmetry. Then, let us pick up such a group now
denoted as $G_0$ in a self-contragredient IR $\omega_0$. Write the
eq.(4) now as
\begin{equation}
\omega_0 (x)=-S_0^{-1}[\omega_0 (x)]^TS_0~,
\end{equation}
for an inner automorphism with $S_0\in\omega_0(G_0)$ for the $\omega_0(G_0)$.
Then, from the theorem 3, the $G_0$ in the IR $\omega_0$ is either a
subgroup of $SO(dim(\omega_0))$ or $SP(dim(\omega_0))$ in the fundamental
vector representation corresponding to a symmetric or skew-symmetric $S_0$.
It is a known fact that $SU(N)$ contains a subgroup SO(N) and
also a subgroup $SP(N)$ for N=even. Therefore, in either case of orthogonal
or symplectic $\omega_0 (G_0)$, we can embed the $G_0$ in $\omega_0$ into the
group $G=SU(\dim(\omega_0)))$ in a fundamental representation $\omega_1$.
Where we have used the obvious branching rule that the fundamental
representation $\omega_1$ of $SU(N)$ reduces to the fundamental representation
of dimension N when the $SU(N)$ is restricted to $SO(N)$ or $SP(N)~(N=even)$.
With such an embedding of $G_0$ into the
$G=SU(dim(\omega_0))$, the element $S_0\in\omega_0(G_0)$ is embedded into
an element $S\in\omega_1(SU(dim(\omega_0)))$ such that the S reduces to the
$S_0$ when the G is restricted to the $G_0$. The inner automorphism
with $S_0$ as in eq.(26) for the $\omega_0$ exchanges the fermions
with their corresponding antifermions as the $\omega_0$ is
self-contragredient, then the fermions in each fermion-antifermion pair in the
$\omega_0$ will be exchanged by the inner automorphism with the S
as in eq.(4) for the $\omega_1(G)$ since the inner automorphism with $S_0$
in the $\omega_0$ is induced by the inner automorphism with the S in the
$\omega_1$ upon the restriction of the G to the $G_0$.
We note here that we only require some of the fermion-antifermion pairs
in the $\omega_0$ to correspond to fermion-antifermion pairs at low energies,
an example is the fermion content in superstring theory with $E_8\times E_8$.

As one can see clearly now, we may choose the unification Yang-Mills
group with CP being a gauge symmetry as $G_{YM}=G=SU(dim(\omega_0)))$ in the
fundamental IR which is complex. Therefore, for each possibility of $G_{YM}$
in the section 2, we can construct another possibility of $G_{YM}$ as a special
unitary group in a fundamental IR. We note that a fundamental vector
representation of $SO(N)$ and $SP(N)~(N=even)$ are always self-contragredient
(actually orthogonal and symplectic for the $SO(N)$ and $SP(N)$ respectively).
In particular for $SO(4n+2)$, the fundamental vector representation
with the highest weight $\Lambda=\lambda_1$ satisfies the self-contragredient
condition eq.(24) with $m_{2n}=m_{2n+1}=0$.

Therefore, we have seen that {\it the $SU(N)$ in the fundamental representation
can be possibilities for the unification Yang-Mills group
$G_{YM}$ with CP as a gauge symmetry}. Some remarks are now in order.

\underline{{\it Remark 3}}. When $G_0$ in a self-contragredient $\omega_0$ is
a possibility for the $G_{YM}$ with CP as a gauge symmetry, then the
$G=SU(dim(\omega_0))$ in the fundamental representation $\omega_1$
as a possibility for the $G_{YM}$ has a larger gauge symmetry than the $G_0$,
since the $SO(dim(\omega_0))$ or $SP(dim(\omega_0))$, for $dim(\omega_0)=even$
in the fundamental vector representation is a non-trivial subgroup of the
$SU(dim(\omega_0))$. However, since $dim(\omega_1)=dim(\omega_0)$ by our
construction, they have the same number of Weyl fermions, and they can have
the same fermion spectrum at the low energies. There exist differences at
the Planck scale for the fermions in $\omega_0(G_0)$ and $\omega_1(G)$.
Although the inner automorphism with the $S_0$ in $G_0$ as in eq.(26)
reverse the sign for all the Yang-Mills charges of the particles in
the $\omega_0(G_0)$, but the corresponding
particles in the $\omega_(G)$ have more Yang-Mills charges,
the inner automorphism with S in the $G$ as in eq.(4) can only
reverse the sign for the Yang-Mills charges contained in the subgroup
$G_0$ embedded into the G, but not for all of them in the $G$ since the
fundamental representation $\omega_1(SU(dim(\omega_0)))$ is complex.
Therefore, our construction above for the $G_{YM}=SU(N)$ is in
the possibilities of type (1) clarified earlier.
A fermion pair corresponding to a fermion-antifermion pair at low energies
may not be a particle-antiparticle pair in the $\omega_1$ for the $G_{YM}$
at the Planck scale.

\underline{{\it Remark 4}}.
Since the inner automorphism (denoted by $X_{YM} $with S as eq.(4) for
the $G_{YM}$ is more non-trivial than a sign-reversing for all the
Yang-Mills charges, it is an inner automorphism of order higher than two.
{\it Therefore, the $CP=X_L X_g X_{YM}$ with $(X_{YM}=S)$ in this case
is a physics operator as an inner automorphism of order higher than two}
Of course, at low energies, it still behaves as a transformation of order two.

\underline{{\it Remark 5}}.
For a self-contragredient IR $\omega_0(G_0)$ of even dimensions,
the $G_0$ in the $\omega_0$ is then
a subgroup of either $SO(dim(\omega_0))$ or $SP(dim(\omega_0))$ in the
fundamental vector IR but not both, due to the fact that a self-contragredient
IR of a semisimple Lie group must be either orthogonal or symplectic
but not both [20,23], and the fundamental vector IR's of $SO(N)$ and $SP(N)$
are orthogonal and symplectic respectively. Therefore, although both
an orthogonal $\omega_0$ and a symplectic $\omega_0$ of even dimensions
can be embedded into the $G_{YM}=SU(dim(\omega_0))$ in a fundamental IR,
only one of the ways (either through orthogonal and symplectic group)
of construction of the $G_{YM}$ can be implemented for a given
self-contragredient $\omega_0$.

\underline{{\it Remark 6}}.
The $SU(N)$ group has two inequivalent fundamental
representations which are contragredient to each other. If a
$\omega_0 (G_0)$ can be embedded into a fundamental IR $\omega_1$ of
$SU(N)$, then the contragredient IR of the $\omega_0$ can be embedded
into the contragredient IR of $\omega_1$ of the $SU(N)$. Since
$\omega_0(G_0)$ is self-contragredient, it can be embedded
into either one of the fundamental IR's of the $SU(N)$, the $SU(N)$
in the two fundamental IR's are both possibilities for the $G_{YM}$.

\underline{{\it Remark 7}}.
As an explicit example, the $E_8$ in the adjoint
(fundamental) representation of dimension 248 is an orthogonal (real)
representation (it is known that the representations of $E_8$ are all
orthogonal). As in superstring theory [14], all the low energy fermions
and their antifermions may be contained in this IR. If we pick up this
$E_8$ as the $G_0$, according to the above construction, we arrive at
the unification Yang-Mills group $G_{YM}=SU(248)$ in the fundamental
representation. (For the embedding of $E_8\times E_8$ in the IR (248,248),
one may have $G_{YM}=SU(248)\times E_8, SU(248)\times SU(248)$ in the
IR (248,248), or $G_{YM}=SU(248\times 248)$ in the fundamental IR of dimension
$248^2=61504$.) Another example is $G_0=SO(32)$ in
the adjoint IR of dimension 496, in this case, one obtains $G_{YM}=SU(496)$
in its fundamental IR.

We note that self-contragredient IR's have interesting relevance to
both local (perturbative) and global (non-perturbative) gauge
anomalies [26-34].
The fact that a self-contragredient IR for a simple group is
a subgroup of some $SO(N)$ or some $SP(N)$ has also been used for
the study of global gauge anomalies [32]. In the present paper, we have used
such fact which is true generally for semisimple Lie groups [20,23].
We have also used the embedding of the fundamental vector IR of $SO(N)$ or
$SP(N)$, and its inner automorphism for the self-contragrediency
into the fundamental IR (complex) of $SU(N)$, and an inner automorphism.
Especially, our embedding of the inner automorphism leads to the
physics operator $CP$ as an inner automorphism of order higher than two.

As we have seen that with a self-contragredient IR $\omega_0$
for some $G_0$ as a possibility for the $G_{YM}$, we may construct
another possibility $G_{YM}=SU(dim(\omega_0))$ in a fundamental IR $\omega_1$.
Actually, this may be further generalized to the following.
{\it When an $SU(N_1)$ in a fundamental IR $\omega_1$ is a possibility
for the $G_{YM}$ with CP as a gauge symmetry, then $SU(N),for~any~N>N_1$
in a fundamental IR $\omega$ may all be a possibility for the $G_{YM}$}.
This may be constructed by embedding the $SU(N_1)$ in a fundamental IR
$\omega_1$ into a fundamental IR $\omega$ of the $SU(N)$.
The embedding also leads to an inner automorphism $X_{YM}$ in the
$SU(N)$ for the $CP=X_L X_g X_{YM}$ corresponding to the $G_{YM}=SU(N)$.
The CP in this case also arises as an inner automorphism of higher
than two for the local symmetry group. However, to be convincing,
we must show that the larger $SU(N)$ may lead to the same low
energy fermions. In other words, the additional fermions need to obtain
their masses at higher energies. To see that this is implemented, let us
note that there is a branching rule that the fundamental IR $\omega$ of
$SU(N)$ reduces to the a fundamental IR $\omega_1$ of $SU(N'),N'<N$
plus $(N-N')$ singlets of $SU(N')$. This or special cases of it
was also used for the other studies [27-34].
One can see this branching rule
obviously by noting that the fundamental IR of $SU(N)$ for any $N\ge3$
reduces to the a fundamental IR of $SU(N-1)$ plus a singlet when $SU(N)$ is
restricted to the $SU(N-1)$. Therefore, symmetry breaking mechanism
(such as spontaneous symmetry breaking) may be implemented to break the gauge
symmetry from the $SU(N)$ in the $\omega$ to the $SU(N_1)$ in the $\omega_1$
(or to the $G_0$ in a self-contragredient $\omega_0$; $SO(dim(\omega_0))$
(for orthogonal $\omega$) or $SP(dim(\omega_0))$ (for symplectic $\omega_0$)
in the fundamental vector IR), such a symmetry breaking may be realized in
one or more steps since the branching rule applies to any pair of $N'<N$.
The consequence is that all the additional fermions are the singlets at
some higher energies. According to the survival hypothesis [35], if
a unification group G breaks at superlarge mass scales down to
$SU(3)\times SU(2)\times U(1)$, then all fermions which can have G-invariant
masses will naturally have such masses (unless some unbroken symmetry prevents
it), the additional fermions in the $\omega$ for the $SU(N)$ in our
construction will obtain their masses at superlarge scales
since a fermion as a G singlet can automatically have G-invariant mass
if no unbroken symmetry prevents it. The theory with $G_{YM}=SU(N)$
constructed in this way may be in general different from that obtained
earlier by embedding a self-contragredient IR $\omega_0$ of $G_0$
into the fundamental vector IR of either $SO(dim(\omega_0))$ or
$SP(dim(\omega_0))$ and then into the fundamental IR $\omega_1$
of $SU(N)$ for $N=dim(\omega_0)$.

\underline{{\it Remark 8}}. We have seen again that in general,
{\it there may be more than one possible theory with $G_{YM}=SU(N)$
(N large enough to have a consistent spectrum of low energy fermions)
in a fundamental IR with CP arising as a gauge symmetry}.

Our result can be easily generalized to include the cases of $SU(N)$ in
the other basic IR's as the possibilities for the $G_{YM}$. This generalization
is due to the branching rules for the basic IR's of $SU(N)$ to $SO(N)$ as
shown in the appendix.

Let us denote an IR with the highest weight $\Lambda$ as $\omega_{\Lambda}$,
and denote the $\omega_{\lambda_i}$ corresponding to a fundamental
weight $\lambda_i$ simply as $\omega_i$ (i=1,2,...,r=rank) as in the appendix.
According to the theorem A1 in the appendix,
the basic IR's $\omega_k$ and $\omega_{N-K}$
of $SU(N)$ reduce to the corresponding IR $\omega_k$ of $SO(N)$ when $SU(N)$
is restricted to $SO(N)$, where
$k=1,2,...,\frac{N-3}{2}~for~N=odd,~and~
k=1,2,...,\frac{N}{2}-2~for~N=even\ge 10$.
Furthermore, these relevant
basic IR's for $SO(N)$ in this case are all self-contragredient (see appendix).
The $SO(N)$ in these IR's could be $G_{YM}$ as we have
seen before, then with the embedding of the $SO(N)$ in these basic IR's into
the corresponding basic IR's (or their contragredient basic IR's), we can see
that $SU(N)$ in these basic IR's could also be
$G_{YM}$ with CP as a gauge symmetry.

Now one can easily see that
the basic IR's listed above covered all the complex basic IR's of $SU(N)$
except that $\omega_{N/2-1}$ and its contragredient IR $\omega_{N/2+1}$ when
$N=even$. The IR $\omega_{N/2}$ ($N=even$) of $SU(N)$ is self-contragredient
and its possibility for the $G_{YM}$ has been discussed in the Section 2.
According to the theorem A2 shown in the appendix,
the $\omega_{N/2-1}$ and $\omega_{N/2+1}$ ($N=even$) of $SU(N)$
will both reduce to the IR
$\omega_{\lambda_{N/2-1}+\lambda_{N/2+1}}=\overline{\rho_1\times\rho_2}$
with the highest weight $\lambda_{N/2-1}+\lambda_{N/2+1}$ of $SO(N)$
when $SU(N)$ is restricted to $SO(N)$, where the
$\overline{\rho_1\times\rho_2}$ denotes the Cartan composition (see appendix)
of the two fundamental spinor IR's $\rho_1$ and $\rho_2$ of the $SO(N)$
($N=even$), i.e. the highest component of the $\rho_1\times\rho_2$.
This IR is obviously self-contragredient and this is true even for $N=4n+2$
since $m_{N/2-1}=m_{N/2+1}=1$ as indicated also in the appendix.
Therefore, by embedding this self-contragredient IR of $SO(N)$
into the basic IR $\omega_{N/2-1}$ or $\omega_{N/2+1}$ of $SU(N)$
($N=even$), we can then see that the complex basic IR's $\omega_{N/2-1}$ and
$\omega_{N/2+1}$ could also be $G_{YM}$. We have therefore
reached the following conclusion.

\underline{{\it Remark 9}}.
{\it $SU(N)$ in the basic IR's $\omega_k$ with the highest weight $\lambda_k$
$(k=1,2,...,N-1)$ could all be the unification Yang-Mills group
$G_{YM}$ with CP as a gauge symmetry}, where we may assume $N\ge 10$ for
$N=even$.

We note here that the fundamental IR's are the special case of
$k=1$ for $SU(N)$. By embedding the group element $S_0$ for the
inner automorphism in these self-contragredient IR's of $SO(N)$ as in eq.(26)
into the corresponding basic IR's (or their contragredient IR's), the
inner automorphism $X_{YM}$ in the $SU(N)$ can be constructed. For the
basic IR's which are complex ($k\ne \frac{N}{2}~for~N=even$),
the $X_{YM}$ or the $CP=X_L X_g X_{YM}$ corresponds to an inner automorphism
of order higher than two.

\underline{{\it Remark 10}}. As an explicit example, by embedding the
adjoint IR of $SO(32)$ which is a basic IR with the highest weight
$\lambda_2$ and of dimension 496 into the corresponding basic IR
$\omega_2$ (or its contragredient IR $\omega_{30}$) of the highest weight
$\lambda_2$ (or $\lambda_{30}$) for $SU(32)$,
we obtain the $SU(32)$ in the complex basic IR $\omega_2$ (or $\omega_{30}$)
of dimension 496 as a possibility for the unification Yang-Mills group
$G_{YM}$ with CP as a gauge symmetry. We note that the $SO(32)$ in
the adjoint IR is also an interesting Yang-Mills group in superstring theory,
and is known to be a group to have a good spectrum of low energy fermions.

\begin{center}
{\bf II. $\bf{G_{YM}=SO(4n+2)}$}
\end{center}

We will now consider the $SO(4n+2)$ in complex representations as possibilities
of $G_{YM}$. We have seen in eq.(24) that an IR of $SO(4n+2)$ with
the highest weight $\Lambda=\Sigma m_i\lambda_i~(i=1,2,...,r=2n+1)$ is
complex if $m_{2n}\ne m_{2n+1}$. It is known that [36] the IR can also
be characterized by the $f_j$'s defined by
\begin{eqnarray}
f_j=m_j+m_{j+1}+...+m_{2n-1}+\frac{1}{2}(m_{2n}+m_{2n+1}),1\le j\le 2n-1,\\
f_{2n}=\frac{1}{2}(m_{2n}+m_{2n+1}), f_{2n+1}=\frac{1}{2}(-m_{2n}+m_{2n+1}),
\end{eqnarray}
with
\begin{equation}
f_1\ge f_2\ge ...\ge f_{2n}\ge\mid f_{2n+1}\mid ~.
\end{equation}
Since the $m_j$'s are all non-negative integers, the $f_j$'s assume values
simultaneously either as all integers or all half-integers, which are known
to correspond to tensor or spinor IR's. We can see then that
an IR of $SO(4n+2)$ is complex if and only if $m_{2n}+m_{2n+1}=odd$,
in particular, all the spinor IR's of $SO(4n+2)$ are complex.
Complex tensor IR's are those corresponding to $m_{2n}\ne m_{2n+1}$ and
$m_{2n}$ and $m_{2n+1}$ being both either even or odd.
In the present paper, our specific IR's for the $SO(4n+2)$ as
a possibility for the $G_{YM}$ with CP a gauge symmetry
will be the fundamental spinor IR's.

Our consideration is due to the following fact: The $SO(4n+1)$ has one
fundamental spinor IR of dimension $2^{2n}$
(will be denoted by $\rho_0$ hereafter), which is self-contragredient
with $\Lambda=\lambda_{2n}$ corresponding to the shorter root.
The $SO(4n+2)$ has two fundamental spinor IR's (will be denoted by $\rho$
and $\bar{\rho}$) also of dimension $2^{2n}$,
which are contragredient to each other. They correspond
to $\Lambda=\lambda_{2n},\lambda_{2n+1}$ respectively. The $\rho$ or
$\bar{\rho}$ reduces to the $\rho_0$ when the $SO(4n+2)$ is restricted to
the $SO(4n+1)$, i.e. the $\rho_0$ can be embedded into either $\rho$ or
$\bar{\rho}$ since it is self-contragredient. This fact was also used [28]
for the study of $SO(N)$ global gauge anomalies.
As we have seen in Section 2,
$SO(4n+1)$ (n large enough) may be possible groups as the $G_{YM}$.
Then, similar to the case of $SU(N)$ in the fundamental IR, we may construct
the $SO(4n+1)$ in a fundamental IR as a possible group as the $G_{YM}$ with
CP being a gauge symmetry. Let the $S_0$ be the group element corresponding to
the inner automorphism as in eq.(26) for the $\rho_0$, then its embedding
into the $\rho$ or $\bar{\rho}$ for the $SO(4n+2)$ automatically leads to
a group element $S\in \rho (SO(4n+2))$ (or $\bar{\rho} (SO(4n+2))$.
Similar to the case of $SU(N)$ in the fundamental IR, the inner automorphism
for the $SO(4n+2)$ with the group element $S$ exchange the fermions in
each pair in $\rho$ (or $\bar{\rho}$) corresponding to a fermion-antifermion
pair at low energies. Let the inner automorphism corresponding to the group
element $S$ be $X_{YM}$, then the $CP=X_L X_g X_{YM}$ is a local symmetry
corresponding to the low energy CP operator.

We have seen that {\it the $SO(4n+2)$ in its
fundamental spinor IR's (complex) could also be the
Yang-Mills group $G_{YM}$ with CP arising as a gauge symmetry.
The CP operator at the Planck scale in this case is again an
inner automorphism of order higher than two for the local symmetry group}.

Our conclusions for Sections 2, and 3 can now be summarized as
the following remark:

\underline{{\it Remark 11}}.
The unification Yang-Mills group $G_{YM}$ with CP arising as a
gauge symmetry may be a simple (or semisimple) group. The $SU(N)$
(or $SU(N)$ ideal) may be in self-contragredient IR's and basic IR's,
$SO(4n+2)$ (or $SO(4n+2)$ ideal) may be in self-contragredient
IR's and fundamental spinor IR's, and $E_6$ may be in
self-contragredient IR's. The self-contragredient IR's of
$SU(N)$, $SO(4n+2)$ and $E_6$ are given by conditions eq.(23-25). For the
complex IR's, the $CP=X_L X_g X_{YM}$ (or $X_{YM}$) in general is an
inner automorphism of order higher than two.

\section{Non-semisimple Lie groups as $G_{YM}$,
Relevance to Superstring Theory, and Cosmic Strings}

Our result in the previous sections 2,3 may be generalized to the
non-semisimple case, i.e. the $G_{YM}$ is a product of a semisimple
Lie group and $U(1)'s$. To include $U(1)'s$ in $G_{YM}$, we need
to generalize the $CP=X_L X_g X_{YM}$ to include the transformation
on $U(1)$ charges. There are following two cases.

(1). Low energy fermions do not carry charges for the $U(1)'s$ in the
unification $G_{YM}$ at its scale.
In this case, the $X_{YM}$ may be either the inner automorphism
for the simple or semisimple ideal as we discussed in previous sections or
with an additional factor $X_a$ which reverse the sign for all the abelian
charges also. They are both consistent with the observation that the $CP$
at low energies transform a Weyl fermion into its antifermion since the charges
for the $U(1)'s$ in the $G_{YM}$ decouple to low energy fermions.

(2). Low energy fermions also carry the charges for the $U(1)$ ideals
or some of them in the unification $G_{YM}$. In this case, we may write
$X_{YM}=X_{s}X_a$, where the $X_{s}$ is the inner automorphism for
the simple or semisimple ideal, i.e. the $X_s$ here may be the same as the
$X_{YM}$ for semisimple $G_{YM}$ as in the previous sections, the $X_a$ is
the operation reversing the abelian charges for all the $U(1)$ ideals in
the $G_{YM}$ or at least for those couple to the low energy fermions at the
$G_{YM}$ scale. For those $U(1)$ ideals, they need to be all
spontaneously broken at some heavy scales in order to allow CP to be a gauge
symmetry, since at low energies in this case, those $U(1)'s$ will decouple to
the observed fermions by decoupling theorem [37].
Otherwise if charges for those $U(1)'s$ are observable at low energies,
then similar to the case of a simple group, a low energy Weyl fermion
and its complex conjugate need to be in the same IR of the $U(1)'s$
at the $G_{YM}$ scale in order for CP to be a local symmetry, i.e.
they need to have the same charges for those $U(1)'s$. This implies
that low energy fermions can only have vanishing charges for those
$U(1)'s$ and that is the case (2).

\underline{{\it Remark 12}}.
Therefore, in general, we have seen that there may be $U(1)$
ideals in the unification $G_{YM}$ with CP arising as a gauge symmetry. The
$U(1)'s$ in the $G_{YM}$ need to be broken at high energy scales if fermions
in the low-energy spectrum couple to those $U(1)'s$ at the $G_{YM}$ scale.

{}From the analysis above about the possible $U(1)$ ideals in the $G_{YM}$,
we can also easily see the implications as given by the following remark.

\underline{{\it Remark 13}}. Since the $U_Y(1)$ weak hypercharge couple to
the low energy fermions, then the $U_Y(1)$ cannot be a symmetry generated as
a combination containing any $U(1)$ from the unification $G_{YM}$ scale if
CP arises as a gauge symmetry, i.e. {\it the $U_Y(1)$ for the weak hypercharge
can only arise as a gauge symmetry from the semisimple ideal of
the unification $G_{YM}$ if CP is to be arising as a gauge symmetry.
Therefore, in general the $U_Y(1)$ is traceless in each generation of
quarks and leptons in the standard model}, even if there are $U(1)$ ideals
in the unification $G_{YM}$. We note here that the traceless of $U_Y(1)$
has been often regarded as the vanishing
of a potential mixed gauge-gravitational anomaly (in a triangle diagram with
one external hypercharge generator and two external gravitons) or as an
indirect evidence for grand unification with a simple or semisimple GUT group
since a generator of a semisimple Lie group must be traceless. It seems that
our analysis on CP arising as a gauge symmetry has theoretically shedded a new
light on this observation.

After the various generalizations from Lie groups in eq.(13) or a product of
them which only allow self-contragredient IR's to non-semisimple unification
$G_{YM}$ including complex IR's for the simple or semisimple ideal as well as
possible $U(1)$ ideals with CP arising as a gauge symmetry,
we will now give some brief discussions including anomaly cancellation.
As we have stated earlier that we do not intend to discuss about anomalies
in details, and we assume that at least some of our more generalized
possibilities for the $G_{YM}$ would correspond to anomaly-free theories
in more relax or generalized conditions for anomaly cancellation or
with deeper development of unified model building.

First of all, since our consideration of CP arising as a gauge symmetry
needs to be in higher dimensions [9], the $G_{YM}$ and anomaly cancellation
may be subtle and there may be many possible theories since we can only probe
directly the physics at some low energy scales. The investigations on various
possibilities consistent with certain conditions (with CP arising as a gauge
symmetry in the present discussion) are physically meaningful. Moreover,
as initiated  from the investigation in Kaluza-Klein theories, it was noted
[38] that to have a chiral fermion spectrum in four dimensions, the
$G_{YM}$ must be nontrivial. Different theories may have different mechanisms
for realizing the unification $G_{YM}$, and therefore anomaly cancellations
in general may also depend on the subtleties of the theories.
In the following, we will provide a brief discussion about the relevance of
our generalized consideration with CP as a gauge symmetry to the superstring
theories, as well as a related result for global (non-perturbative) gauge
anomalies.

In superstring theory, it is known that $SO(32)$ and $E_8\times E_8$ for
the $G_{YM}$ in the adjoint IR can correspond to anomaly-free theory in
ten dimensions with Green and Schwarz mechanism [39]. It is known that [39]
in the popular ten-dimensional superstring theory, the anomaly cancellation
in Green Schwarz mechanism leads to the trace identity for Yang-Mills
group of n=496 Yang-Mills symmetries (generators) in the adjoint IR given by
\begin{equation}
TrF^6=\frac{1}{48}TrF^2TrF^4-\frac{1}{14400}(TrF^2)^3~~,
\end{equation}
where the F denotes the Yang-Mills two-forms for the field strength.
For Yang-Mills groups of n=496 generators, this trace identity is
necessary and sufficient for the factorization of the 12-forms
related to the anomalies [39] so that the anomalies can be canceled with
the non-trivial transformation of the antisymmetric tensor B field.
The above trace identity is satisfied by $SO(32)$, and $E_8\times E_8$
which leads to the heterotic string theory in ten dimensions.
The heterotic string theories with the Yang-Mills group $SO(32)$ and
$E_8\times E_8$ are known to be also free of global (non-perturbative)
gauge anomaly due to the fact that the relevant homotopy group
$\Pi_{10}(SO(32))=\Pi_{10}(E_8\times E_8)=\{0\}$ is trivial.
Since both $SO(32)$ and $E_8\times E_8$ only allow self-contragredient
IR's, the heterotic string theory with these two Yang-Mills groups may
consistently allow CP to arise as a gauge symmetry [9]. The ref.18-19
for local anomaly theories have used more relaxed conditions including also
the factorization similar to that in Green-Schwarz mechanism. However,
we will not go into details here.

Since self-contragredient IR's are interesting in considering the
possible Yang-Mills groups with CP arising as a gauge symmetry in
8k+2 dimensions, we would like to include in the following a result in ref.27,
although we will not discuss about the anomaly cancellation more generally.
{\it If we require the strong anomaly-cancellation condition
$Tr^{(\omega)}F^{4k+2}=0$ for a Yang-Mills group in a representation $\omega$
for local (perturbative) gauge anomalies, then any self-contragredient $\omega$
is also free of global gauge anomalies in the 8k+2 dimensions}.
This is due to the fact that the global gauge anomaly coefficient with
strong anomaly-cancellation condition for local gauge anomalies will be
given by
\begin{equation}
A(\omega)=(-1)^{indD_{8k+2}}=1,
\end{equation}
since the Dirac index consists of odd-order traces of F,
which vanishes for a self-contragredient representation.
Actually, this result is true generally in 4k+2 dimensions although the 8k+2
dimensions are more relevant to our consideration of CP arising as a gauge
symmetry [9].

In fermionic string models with a fermionic formulation of all internal
(i.e. toroidally compactified) coordinates, the fermionized internal
coordinates can be treated as world sheet fermions which are specified by their
boundary conditions and by their interactions on the world sheet.
It is known that (for a review, see ref.40), many string models with different
Yang-Mills groups may be consistently constructed in $D\le 10$ dimensions.
As it is discussed in ref.40 that in the first quantized formalism,
all consistent string models should be treated on an equal footing, although
the ten-dimensional $E_8\times E_8$ and $SO(32)$ heterotic strings correspond
to more obvious constructions.
Then [40] it is plausible that the string dynamics may select
a subset of the first quantized string models as locally stable states
(or second quantized vacuum states), and our universe may be sitting at a
locally stable point in string field space (not necessarily the unique ground
state the string dynamics may eventually select, since [40] the tunneling from
some locally stable state to this possible unique ground state typically
may take a time many orders of magnitude longer than the age of the universe
due to the Planck scale involved). Therefore all the consistent string models
are physically interesting from this point of view.  Related to our
generalized consideration of CP arising as a gauge symmetry, we note that
some consistent fermionic string models can have groups of the form $SO(4n+2)$
or $SU(N)$ (e.g. $SO(14)$, $SO(10)$, and $SU(8)$ in complex IR's) as an ideal
in the Yang-Mills group [40]. For these Yang-Mills groups, even if some
ideals (e.g. $E_7$ in $SO(4)\times U(8)\times E_7\times SO(8)\times SU(2)$
which is an explicit example in ref.40) only have self-contragredient IR's
which is consistent with CP as a gauge symmetry as in ref.9,
the relevant representations for the Yang-Mills group in the model
may be complex (in fact containing $U(1)$ ideal in the example). Therefore, our
generalized construction of inner automorphism for complex IR's
by embedding method for the CP
and the inclusion of $U(1)$ ideals may be of interest in the investigation of
consistent string models. We note here also that we have been only using
examples to clarify the possible relevance of our generalized consideration
with CP as a gauge symmetry to string models. For some string models,
our generalized construction with CP as a gauge symmetry will be
rather involved.

Obviously, if the unification Yang-Mills group contains $U(1)$ ideals, then
it may have effect on cosmic strings.  Topologically stable strings in
four dimensions are possible whenever a simply connected internal gauge group
$G$ breaks to a group $H$ such that the $H$ has disconnected components or
the manifold $G/H$ is not simply connected.
As discussed in ref.9, when the vacuum configuration spontaneously breaks $G$
to
$H$ which includes CP as a discrete element, i.e. CP is an unbroken element
of a spontaneously broken continuous symmetry, then stable CP strings should
exit. The discussions on CP strings etc. in ref.9 may also apply in our
generalized consideration when the Yang-Mills group is simply connected.
In particular, it may have effect on CP violations at low energies.
When the unification Yang-Mills group contains $U(1)$ ideals, then the
CP strings may not exist when the manifold $G/H$ is simply connected, where
note that the fundamental group $\Pi_1(G/H)$ may be trivial if
$G$ is not simply connected as one can see from the relevant exact homotopy
sequence [41].
\section{Conclusions}
In this paper, we have generalized the consideration of possible candidates
for the unification Yang-Mills group $G_{YM}$ with CP arising as a gauge
symmetry, which is initially discussed in ref.9. Our generalization is
significant due to its possible role in unified model building, as well as
its connection to CP violation, especially the solution to the strong
CP problem in the models with CP spontaneously broken [10]. Our
generalization with CP as a gauge symmetry is interesting due to
the argument [12-14] that global symmetries in effective theories below
the Planck scale may be violated by quantum gravity effects from wormholes
and virtual black holes or non-perturbative effects in string theory.

Besides the groups which only allow self-contragredient representations,
i.e. $E_8$, $E_7$, $SO(2n+1)$, $SO(4n)$, $SP(2n)$,
$G_2$ or $F_4$ (or a product of them) for the $G_{YM}$ with CP as a gauge
symmetry as in the discussions of ref.9, we have found many other
possible groups as the $G_{YM}$ in our generalized consideration in this paper.
As we have seen in Sections 2,3 in our generalized consideration,
the unification Yang-Mills group $G_{YM}$ with CP arising as a
gauge symmetry may be a rather general simple (or semisimple) group,
including also $SU(N)$, $SO(4n+2)$ and $E_6$. The $SU(N)$
(or $SU(N)$ ideal) may be in self-contragredient IR's and basic IR's
(including the fundamental IR's, see appendix),
$SO(4n+2)$ (or $SO(4n+2)$ ideal) may be in self-contragredient
IR's and fundamental spinor IR's, and $E_6$ may be in
self-contragredient IR's. The self-contragredient IR's of $SU(N)$,
$SO(4n+2)$ and $E_6$ are given by conditions eq.(23-25).
We note here that many IR's listed above are complex.
For the complex IR's, the $CP=X_L X_g X_{YM}$ (or $X_{YM}$) in general is an
inner automorphism of order higher than two. We have shown that how such an
inner automorphism of order higher than two for a complex IR may be constructed
through the embedding method. Such an inner automorphism reverses the signs
of the Yang-Mills charges corresponding to low-energy gauge symmetries for
the fermions in the low-energy spectra, and this is consistent with the
observed transformation properties of CP at low energies.
To the best knowledge of the present author, such a construction of
an inner automorphism of order higher than two for a physics operator
has never been given before [25].

We have also generalized our consideration to include non-semisimple
groups as the unification $G_{YM}$ with CP as a gauge symmetry.
An interesting finding is that with CP as a gauge symmetry, the
weak hypercharge $U_Y(1)$ in the standard model needs to be generally
traceless in each generation of quarks and leptons even if the
$G_{YM}$ is not semisimple, i.e. it has $U(1)$ ideals. Since all we need
to reach this conclusion is that the $U_Y(1)$ is a low-energy gauge symmetry,
we then know that this conclusion can be theoretically more general, i.e.
for any low-energy $U(1)$ gauge symmetry, the generator needs to be traceless
even if the $G_{YM}$ with CP as a gauge symmetry contains $U(1)$ ideals.
We have also given a brief discussion on the possible relevance to the
superstring theories and cosmic strings, as well as a result in ref.27
for the absence of global gauge anomalies for self-contragredient
representations in 8k+2 dimensions related to the consideration of
CP arising as a gauge symmetry.

We like to indicate here that although many groups could be the unification
Yang-Mills group $G_{YM}$ theoretically,
{\it unification Yang-Mills groups with only unitary ideals
seem more natural from the consideration that the observed gauge theories
at low energies are all of unitary groups}. Our generalized
consideration of including $SU(N)$ and $U(1)$ ideals in the unification
$G_{YM}$ with CP arising as a gauge symmetry could be of interest from this
point of view although we have not seen before such a support for the unitary
groups (besides that it can have complex representations) for unified model
building by the unitary form of the gauge groups at low energies.

Moreover, we have also given an appendix in which we show a theorem on
the branching rules for the basic representations of $SU(N)$ to $SO(N)$.
We expect that our generalized consideration of unification Yang-Mills
groups and representations with CP arising as a gauge symmetry
and construction of relevant inner automorphisms by embedding may be
useful in the general study of non-abelian gauge theories as well as
in the unification theory.

\underline{Acknowledgement}: I would like to express my gratitude to
Geoffrey West and the Particle Theory Group at Los Alamos National Lab
for hospitality. This work is supported by a DOE fellowship.
\newpage
{\bf\large Appendix: Branching Rules for the Basic Representations of
$SU(N)$ to $SO(N)$}

In this appendix, we will mainly show a theorem for the branching rules
for the basic representations of $SU(N)$ to $SO(N)$ which is used for the
construction of $SU(N)$ in the basic representations (complex except that
with the highest weight $\Lambda=\lambda_{(r+1)/2}$ when the rank $r=N-1=odd$)
as the unification Yang-Mills group $G_{YM}$ with CP arising as a gauge
symmetry. For a review of Lie groups related to gauge symmetry, as well as
spacetime symmetries which are relevant to the construction of CP operator
corresponding to a gauge symmetry, see ref.36, and 42.
To clarify the result we need, it will be useful to recall some
relevant definitions [23], and we will denote an IR of a simple Lie algebra
$\cal G$ with the highest weight $\Lambda$ as $\omega_{\Lambda}$.

\underline{{\it Basic Representations}}:
For a simple Lie algebra $\cal G$, an IR $\omega$
is basic if and only if its highest weight is a fundamental weight,
i.e for the highest weight
$\Lambda=\Sigma m_i\lambda_i$, all the $m_i$'s are zero except one that is 1.

\underline{{\it Fundamental Representations}}:
The basic IR's of a simple Lie algebra
$\cal G$ corresponding to terminal points in the
Dynkin diagram are fundamental IR's, here a terminal point is one
that is connected with not more than one point in the Dynkin diagram.

\underline{{\it Cartan Composition}}:
Let $\omega_{\Lambda}$ and $\tau_{\Lambda'}$ be
two IR's of $\cal G$. Then the tensor (Kronecker) product representation
$\omega\times\tau$ is in general reducible. Its highest component
denoted by $\overline{\omega\times\tau}$
is an irreducible IR with the highest weight $\Lambda + {\Lambda}'$.
The operation of tensor multiplication combined with the operation of
separating the highest component then lead to the formation of a new
IR, this composite operation is called the {\it Cartan composition} of
the IR's. The Cartan composition can be obviously applied to more than two
IR's. Where one needs to distinguish the notation for the
Cartan composition from that for a self-contragredient representation
(with a short bar).

\underline{{\it k-th Alternation}}:
Let $\omega$ be a linear representation of
dimension n for a Lie algebra, by tensor products in its representation
vector space $R_{\omega}$ of dimension n, we can construct a linear
space $R_{\omega}^{\{k\}}$ consisting of all the antisymmetric tensors of
rank k ($k\le n$). This space is then a representation space
of dimension $C^k_n$. The space $R_{\omega}^{\{k\}}$
is called the {\it k-th alternation} of
the $R_{\omega}$, and the corresponding representation
denoted by $\omega^{\{k\}}$ is called the {\it k-th alternation}
of $\omega$. The highest component denoted by $\overline{\omega^{\{k\}}}$
for the k-th alternation of $\omega$ has the highest weight
$\Lambda_1+\Lambda_2+...+\Lambda_k$, where $\Lambda_1,\Lambda_2,...,\Lambda_k$
are the k highest weight of $\omega$, and  the case of k=1 means the $\omega$
itself.

\underline{{\it k-th Symmetrization}}:
Similar to the alternation above, we can also
construct of symmetric tensor space $R_{\omega}^{[k]}$ or $R_{\omega}^{<k>}$
$(k>0)$ of dimension $C^k_{n+k}$,
which is called the {\it k-th symmetrization} of the
linear space $R_{\omega}$, and it is also a representation space
of the Lie algebra. The corresponding representation denoted by
$\omega^{[k]}$ is call the {\it k-th symmetrization} of $\omega$.
The highest component of it will be denoted by $\overline{\omega^{[k]}}$,
obviously it has the highest weight $k\Lambda$ with $\Lambda$ being the
highest weight of $\omega$. Note that the Cartan composition
$\overline{\omega (1)\times\omega(2)\times ... \times\omega (k)}$ with
$\omega (1),\omega (2),...,\omega (k)$ all equivalent to $\omega$ has
the highest weight $k\Lambda$, therefore,
the $\omega^{[k]}$ can be regarded as the
result of a repeated Cartan composition of $\omega$ with itself.

It is known that [23] for a simple Lie algebra, the operation
$\overline{\omega^{\{k\}}}$ and the
Cartan composition enable us to construct arbitrary IR's from
fundamental representations. In particular,
an arbitrary basic IR of a simple Lie algebra
can be obtained as $\overline{\omega^{\{k\}}}$ from some fundamental
representation for some integer $k>0$. The fundamental IR's are
special cases of basic IR's (k=1).

In the present paper, the branching rules needed are for classical
and simple groups, which we will focus on hereafter.
In the following discussions, we will denote the basic IR's
$\omega_{\lambda_i}$ simply as $\omega_i~(i=1,2,...,r=rank)$ for
convenience. The basic IR's in terms of the fundamental IR's are given
as follows.\\
\underline{$A_n=SU(n+1)$}:

Fundamental IR's: $\omega_1,\omega_n$ (which are self-contragredient
to each other).

It is known that for every k=1,2,...,n, ${\omega_1}^{\{k\}}$ and
${\omega_n}^{\{k\}}$ are irreducible, so that
\begin{eqnarray}
\overline{{\omega_1}^{\{k\}}}={\omega_1}^{\{k\}},\\
\overline{{\omega_n}^{\{k\}}}={\omega_n}^{\{k\}}.
\end{eqnarray}
In terms of the fundamental IR's, the basic IR's are given by the
alternations as
\begin{equation}
\omega_k={\omega_1}^{\{k\}}={\omega_n}^{\{n+1-k\}},~k=1,2,...,n.
\end{equation}

\underline{$C_n=SP(2n)$}:

Fundamental IR's: $\omega_1,\omega_n$

The basic IR's can all be obtained as the alternations of
the fundamental IR, i.e.
\begin{eqnarray}
\omega_k=\overline{{\omega_1}^{\{k\}}}~~(k=1,2,...,n),
\end{eqnarray}
where note that n-th simple root has longer length in the Dynkin diagram.

\underline{$B_n=SO(2n+1)$}:

Fundamental vector IR: $\omega_1$

Fundamental spinor IR: $\omega_n=\rho_0$\\
The other basic IR's: $\overline{\omega_1^{\{k\}}}$ (k=1,2,..,n-1),\\
it can be shown that the $\omega_1^{\{k\}}$ are irreducible
for k=1,2,..,n, therefore, those basic IR's can also be written as
\begin{equation}
\omega_k=\overline{\omega_1^{\{k\}}}=\omega_1^{\{k\}}~~(k=1,2,..,n-1).
\end{equation}
The relations to the fundamental spinor IR are given by
\begin{eqnarray}
\rho_0^{<2>}=\omega_1^{\{n\}},
\overline{\rho_0^{\{2\}}}=\omega_1^{\{n-1\}}=\omega_{n-1}.
\end{eqnarray}

\underline{$D_n=SO(2n)$}~$(n\ge 5)$:

Fundamental vector IR: $\omega_1$

Fundamental spinor IR's: $\omega_{n-1}=\rho_1,\omega_n=\rho_2$\\\
The other basic IR's: $\overline{\omega_1^{\{k\}}}$ (k=1,2,..,n-2),\\
the $\omega_1^{\{k\}}$ are irreducible
for k=1,2,..,n-1, therefore, those basic IR's can also be written as
\begin{equation}
\omega_k=\overline{\omega_1^{\{k\}}}=\omega_1^{\{k\}}~~(k=1,2,..,n-2).
\end{equation}
The relations to the fundamental spinor IR's are given by
\begin{eqnarray}
\overline{\rho_1\times\rho_2}=\omega_1^{\{n-1\}},
\rho_1^{<2>}\oplus\rho_2^{<2>}=\omega_1^{\{n\}},
\overline{\rho_1^{\{2\}}}\simeq\overline{\rho_2^{\{2\}}}
\simeq\omega_1^{\{n-2\}}.
\end{eqnarray}

Now we can show the branching rules we need for the $SU(N)$ to $SO(N)$
in basic IR's. $SU(N)$ contains $SO(N)$ as a subgroup. An IR of $SU(N)$
in general will reduce to a direct sum of IR's of $SO(N)$ when the $SU(N)$
is restricted to the $SO(N)$ subgroup.

Let us note first that for the linear space consisting of all the
antisymmetric tensors of rank k constructed from the vector space of
dimension n, its dimension is $C^k_n$. From this, we can see then that
the dimensions for the basic IR's for $SU(N)$ are given by
\begin{equation}
d(\omega_k)=C^k_{N}=\frac{N!}{k!(N-k)!}~~(k=1,2,...,N-1).
\end{equation}
For $SO(N)$, the basic IR's other than the fundamental spinor
IR(s) have the dimensions given similarly by
\begin{equation}
d(\omega_k)=C^k_{N},
\end{equation}
here
\begin{equation}
(k=1,2,...,\frac{N-3}{2}~for~N=odd),
(k=1,2,...,\frac{N}{2}-2~for~N=even\ge 10).
\end{equation}
We note here that these basic IR's for the $SO(N)$ are all
self-contragredient $(m_{N/2}=m_{N/2-1}=0~for~N=even)$, and then
there is no other inequivalent IR's with the same dimensions.
Comparing the dimensions of the basic IR's for the $SU(N)$ and $SO(N)$,
we can then see the branching rules we need. Therefore, we have shown the
following theorem.

\underline{{\it Theorem A1}}.{\it
The basic IR's $\omega_k$ and $\omega_{N-k}$ with the highest weight
$\Lambda=\lambda_k,~and~\lambda_{N-k}$ for
$SU(N)$ reduce to the corresponding basic IR $\omega_k$ with the highest
weight $\lambda_k$ for $SO(N)$ when $SU(N)$ is restricted to $SO(N)$,
where $k=1,2,...,\frac{N-3}{2}~for~N=odd,~and~
k=1,2,...,\frac{N}{2}-2~for~N=even\ge 10$.}

\underline{{\it Remark A1}}. We note that the relevant basic IR's for
$SU(N)$ in the theorem A1 above are all complex.
Since the relevant basic IR's for the $SO(N)$ are all self-contragredient,
the branching rules then also apply to the contragredient IR's of
$\omega_k$ for the $SU(N)$. Here, the
contragredient of $\omega_k$ for $SU(N)$ is equivalent to $\omega_{N-k}$.

We can also see easily that for $N=even\ge 10$, the $\omega_1^{\{N/2-1\}}$
for both $SU(N)$ and $SO(N)$ have the dimension $C^{N/2-1}_N$, and it is
an IR for both of them, therefore, we have also shown the following theorem.

\underline{{\it Theorem A2}}.
{\it The basic IR $\omega_{N/2-1}=\omega_1^{\{N/2-1\}}$ and its
contragredient IR $\omega_{N/2+1}$ ($N\ge 10$)
with the highest weights $\Lambda=\lambda_{N/2-1}$ and $\lambda_{N/2+1}$
of $SU(N)$ reduce to the IR
$\omega_1^{\{N/2-1\}}=\overline{\rho_1\times\rho_2}$
with the highest weight $\Lambda=\lambda_{N/2-1}+\lambda_{N/2}$
when $SU(N)$ is restricted to $SO(N)$}.

\underline{{\it Remark A2}}. In the theorem A2, the
IR $\omega_1^{\{N/2-1\}}=\overline{\rho_1\times\rho_2}$ for $SO(N)$ is
not a basic IR. Since $m_{N/2-1}=m_{N/2}=1$, it is a self-contragredient
IR even if $N=4n+2$, but for the $SU(N)$, the contragredient of
$\omega_{N/2-1}=\omega_1^{\{N/2-1\}}$ is
$\omega_{N/2+1}$, it then corresponds to a complex basic IR.

\underline{{\it Remark A3}}. Our theorems A1 and A2 for the branching
rules for $SU(N)$ to $SO(N)$ covered all the
complex basic IR's of $SU(N)$. The only case which is not included
in the theorems is the basic IR $\omega_{N/2}$ when $N=even$, this
IR is self-contragredient. In this case, the $\omega_1^{\{N/2\}}$
($N=even$) for $SO(N)$ is not irreducible, the branching rule
may be more non-trivial. However, we are mainly interested in the branching
rules for the complex IR's of $SU(N)$ to $SO(N)$ for our discussions in
section 3, the self-contragredient IR's are discussed in section 2.

Therefore, we have seen that {\it for every complex basic IR $\omega_k$
of $SU(N)$, there is a self-contragredient IR $\omega'_k$ of $SO(N)$ such
that the $\omega_k$ reduces to the $\omega'_k$ when $SU(N)$ is restricted
to $SO(N)$. The $\omega'_k$ is the corresponding basic IR for $SO(N)$
except when $k=N/2-1$ or $K=N/2+1$ for $N=even$}.


\newpage
\begin{thebibliography}{99}
\bibitem{thfcnc}For a review, see {\it CP Violations}, ed. C. Jarlskog,
(World Scientific, 1989) and references theirin.
\bibitem{thfcnc}G. 't Hooft, {\it Cargese Summer Institute Lectures}, 1979.
\bibitem{thfcnc}R. D. Peccei, H. Quinn, Phys. Rev. Lett. 38, 1440(1977);
Phys. Rev. D16, 1791(1977).
\bibitem{thfcnc}S. Weinberg, Phys. Rev. Lett. 40, 223(1978); F. Wilczek,
Phys. Rev. Lett. 40, 279(1978).
\bibitem{thfcnc}J. E. Kim, Phys. Rev. Lett. 43, 103(1979); A. Vainshtein and
V. Zakharov, Nucl. Phys. B166(1980); A. P. Zhitnitskii, Sov. J. Nucl. Phys.
B31, 260(1980); M. Dine, W. Fischler and M. Srednicki, Phys. Lett. B104,
199(1981).
\bibitem{thfcnc}H. Zhang, Phys. Lett. B322, 374(1994); Erratum-ibid. B326,
329(1994); Phys. Lett. B322, 382(1994); J. Group Theory Phys. Vol.2, No.1,
55-80(1994); Int. J. Mod. Phys. A9, 5729(1994); see also H. Zhang,
in {\it The Fermilab Meeting-DPF92, Vol. II}, Eds. C. Albright,
P. Kasper, R. Raja and J. Yoh (World Scientific, 1993); H. Zhang,
in {\it Proc. Int. Symp. on Fundamental Aspects of Quantum Theory},
(Columbia, South Carolina, 1992), eds. J Safko et al. (World Scientific, 1993).
\bibitem{thfcnc}S. Samuel, Mod. Phys. Lett. A7, 2007(1992).
\bibitem{thfcnc}S. Okubo and R. E. Marshak, Prog. Theor. Phys. 87, 1059(1992).
\bibitem{thfcnc}K. Choi, D. B. Kaplan, and A. E. Nelson, Nucl. Phys. B391,
515(1992). See also M. Dine, R. G. Leigh, and D. A. MacIntire, Phys. Rev. Lett.
69, 2030(1992).
\bibitem{thfcnc}A. E. Nelson, Phys. Lett. B136, 387(1984); S. M. Barr,
Phys. Rev. Lett. 53, 329(1984); S. M. Barr, Phys. Rev. D30, 1805(1984);
D34, 1567(1986).
\bibitem{thfcnc}M. Dine and N. Serberg, Nucl. Phys. B273, 109(1986), J. M.
Flynn and L. Randall, Nucl. Phys. B293, 731(1987).
\bibitem{thfcnc}S. Hawking, Phys. Lett. B195, 337 (1987); G. V. Lavrelashvili,
V. Rubakov, and P. Tinyakov, JETP Lett.46, 167(1987); S. Giddings and A.
Strominger, Nucl. Phys. B307, 854(1988).
\bibitem{thfcnc}S. Hawking, Comm. Math. Phys. 43, 199(1975).
\bibitem{thfcnc}M. B. Green, J. H. Schwarz and E. Witten, {\it Superstring
Theory, Vol.2}, (Cambridge Univ. Press, 1987), and references therein.
\bibitem{thfcnc}T. Banks, Physicalia, 12, 19(1990).
\bibitem{thfcnc}L. Krauss and F. Wilczek, Phys. Rev. Lett.62, 1221(1989).
\bibitem{thfcnc}J. Preskill and L. Krauss, Nucl. Phys. B341, 50(1990);
M. Alford et al., Nucl. Phys. B337, 695(1990); B349, 414(1991);
Phys. Rev. Lett. 64, 1632, (1990).
\bibitem{thfcnc}J. Thierry-Mieg, Phys. Lett. B171, 163(1986); A. N.
Schellekens, ibid. 175, 41(1986); A. N. Schellekens and N. P. Warner,
Nucl. Phys. B287, 317(1987).
\bibitem{thfcnc} Y. Tosa and S. Okubo, Phys. Rev. D36,
2484(1987); 37, 996(1988).
\bibitem{thfcnc}A. I. Malcev, Am. Math. Soc., Transl., Ser.1, 9, 172(1962);
M. L. Melta, J. Math. Phys. 7, 1824(1966); M. L. Mehta and P. K. Srivastava,
ibid. 7, 1833(1966).
\bibitem{thfcnc}A. K. Bose and J. Patera, J. Math. Phys. 11, 2231(1970).
\bibitem{thfcnc}H. Samelson, {\it Notes on Lie Algebras}, (Van Nostrand
Reinhold, New York, 1969).
\bibitem{thfcnc}E. B. Dynkin, Am. Math. Soc. Transl. Ser.2, 6, 245(1957)
and references therein.
\bibitem{thfcnc}E. B. Dynkin, Dokl. Akad. Nauk SSSR (N.S.)76, 629(1951).
\bibitem{thfcnc}R. Slansky, Phys. Rep. 79, 1(1981).
\bibitem{thfcnc}E. Witten, Phys. Lett. B117,324(1982).
\bibitem{thfcnc}S. Okubo, H. Zhang, Y. Tosa, and R. E. Marshak, Phys. Rev.
D37,1655(1988).
\bibitem{thfcnc}H. Zhang, S. Okubo, and Y. Tosa, Phys. Rev. D37,2946(1988).
\bibitem{thfcnc}See also H. Zhang and S. Okubo, Phys. Rev. D38,1880(1988).
\bibitem{thfcnc}S. Okubo and H. Zhang, in {it Perspectives on Particle
Physics}, ed. by S. Matsuda, T. Muta, and R. Sakaki (World Scientific, 1989).
\bibitem{thfcnc}A. T. Lundell and Y. Tosa, J. Math. Phys. 29, 1795(1988).
\bibitem{thfcnc}S. Okubo and Y. Tosa, Phys. Rev. D40, 1925(1989).
\bibitem{thfcnc}H. Zhang, LBL-35864, 1994; J. Group Theory Phys. Vol.3, No.1,
89(1995).
\bibitem{thfcnc}See also S. Elitzer and V. P. Nair, Nucl. Phys.
B243, 205(1984).
\bibitem{thfcnc}H. Georgi, Nucl. Phys. B156, 126(1977).
\bibitem{thfcnc}H. Weyl, {\it Classical Groups} (Princeton Univ. press,
Princeton, NJ, 1946).
\bibitem{thfcnc}T. Appelquist and J. Carazzone, Phys. Rev. D11, 2856(1975).
\bibitem{thfcnc}M. F. Atiyah, Hirzebruch, in {\it Essays in Topology and
Related Subjects}, et. A. Haefliger and R. Narasimhan (Springer, NY);
E. Witten, Nucl. Phys. B186, 412(1981); C. Wetterich, Nucl. Phys. B223,
109(1983).
\bibitem{thfcnc}M. Green and J. H. Schwarz, Phys. Lett. B149, 117(1985);
Nucl. Phys. B255, 93(1985); M. Green, J. H. Schwarz, and P. West,
Nucl. Phys. B254, 327(1985).
\bibitem{thfcnc}H. Kawai, D. C. Lewellen and S. H. Tye, Nucl. Phys. B288,
1(1987); Phys. Rev. D34, 3749(1986) and references therein; see also
I. Antoniadis, C. Bachas, C. Kounnas, and P. Windey, Phys. Lett.
B171, 51(1986).
\bibitem{thfcnc}S. T. Hu, {\it Homotopy Theory} (Academic, New York, 1959);
N. Steenrod, {\it Topology of Fibre Bundles} (Princeton University,
Princeton, NJ, 1951).
\bibitem{thfcnc}J. Humphrey, {\it Introduction to Lie algebras and
Representation Theory} (Springer-Verlag, New York, 1970); Y.-S. Kim and
M. Noz, {\it Theory and Application of the Poincare Group}
(Reidel Dordrecht, 1986).
\end{thebibliography}
\end{document}